\documentclass[aps,prl,twocolumn,superscriptaddress,showpacs]{revtex4}

\usepackage{graphicx}
\usepackage{amsfonts}
\usepackage{amssymb}
\usepackage[]{amsmath}
\usepackage[]{latexsym}
\usepackage[]{float}
\usepackage{bm}

\begin{document}


\title{Quantum Hall plateau transition in graphene\\ 
with spatially correlated random hopping}

\author{Tohru Kawarabayashi}
\affiliation{Department of Physics, Toho University,
Funabashi, 274-8510 Japan}

\author{Yasuhiro Hatsugai}
\affiliation{Institute of Physics, University of Tsukuba, Tsukuba, 305-8571 Japan}

\author{Hideo Aoki}
\affiliation{Department of Physics, University of Tokyo, Hongo, 
Tokyo 113-0033 Japan }

\date{\today}

\begin{abstract}
We investigate how the criticality of the quantum Hall plateau  
transition in disordered graphene differs from those in the ordinary
quantum Hall systems, 
based on the honeycomb lattice with ripples modeled as random hoppings.
The criticality of the graphene-specific $n=0$ Landau level 
is found to change dramatically 
to an anomalous, almost exact fixed point 
as soon as we make the random hopping spatially correlated over 
a few bond lengths.  
We attribute this to the preserved chiral symmetry 
and suppressed scattering between K and K' points in the Brillouin zone.  
The results suggest that 
a fixed point for random Dirac fermions with chiral symmetry 
can be realized in free-standing, clean graphene with ripples. 
\end{abstract}


\maketitle

After the seminal observation of the anomalous quantum Hall 
effect (QHE) in graphene,\cite{Geim,Kim,ZhengAndo} 
fascination expands with the graphene QHE.  
One crucial question that is not fully explored 
is: what exactly is the role of the chiral symmetry in the 
problem?  
This has to do with a most significant feature of 
double Dirac cones (at K and K' in the Brillouin zone) in graphene. 
Although a single Dirac cone would already 
imply a characteristic Landau level structure 
with the zero-energy level, 
if we really want to look at the effect of disorder on 
the graphene Landau levels, 
we have to go back to the honeycomb lattice for which 
the chiral (A-B sub-lattice) symmetry\cite{HFA,HChiral} and the 
associated valley (K and K') degrees of freedom enter as an 
essential ingredient. The effect of disorder 
should then be sensitive to the nature of disorder, i.e., 
bond disorder or potential disorder, which determines the presence or 
otherwise of the chiral symmetry\cite{KA,SM,OGM}, 
and whether the disorder is short-ranged or long-ranged, which 
controls the scattering between K and K' points. 

For  Dirac fermions, effects of random gauge fields induced by 
ripples in the two-dimensional plane 
have been discussed\cite{Meyer,Neto,Guinea},
and the stability of zero modes has been argued in terms of 
the index theorem and the chiral symmetry\cite{Geim,KN,HFA,HChiral}.
More recently, the plateau-to-plateau transition for random Dirac fermions 
has been discussed, where the particle-hole symmetry 
is shown to 
make the zero-energy Landau level robust 
\cite{NRKMF}.  
As for the criticality, however, the result\cite{NRKMF} 
shows nothing special about 
the $n=0$ Landau level, but this is obtained for 
a model of the Dirac fermions for which the randomness 
is introduced as a scalar random potential, 
so the chiral symmetry is degraded.

On the other hand, the actual randomness in graphene, 
even when atomically clean, 
is known to have ripples, i.e, long-ranged corrugation of 
the graphene plane\cite{Meyer}.  
In fact, while a monolayer graphene naively contradicts 
with the well-known theorem 
that two-dimensional crystals should be thermodynamically  
unstable, one explanation
attributes the stability to the ripples\cite{Fasolino}. 
In this sense, we can take the disorder coming from ripples 
in graphene as an intrinsic disorder.  
Since the ripples consist of random bending 
of the honeycomb lattice, 
its main effect should be, in the tight-binding model, 
a modification of the hopping integral 
between neighboring sites\cite{Neto}. 
Thus the question amounts to: how does the QHE criticality behave for a 
model with {\it random hopping} on the honeycomb lattice.  
The random hopping is of fundamental theoretical interest as well, since 
a bond randomness preserves the chiral symmetry, 
so its effect, particularly on the criticality, is of crucial interest.  
The chiral symmetry indeed plays a fundamental role
in graphene \cite{HFA,HChiral}, which protects the gapless double Dirac cones as well as the 
existence of characteristic zero-modes with/without 
magnetic fields.  
A special importance of the chiral symmetry in localization physics
has also been discussed with a viewpoint of the 
universality\cite{LFSG,HWK,AZ}.

Now, the length scale over which the lattice is warped 
should be reflected as the spatial correlation in the random hopping.  
In the case of the ordinary QHE systems, 
the importance of the spatial correlation of randomness has been discussed 
in various contexts, among which are the pioneering work by 
Ando and Uemura\cite{AndoUemura},  
the levitation of the critical states in the lattice model\cite{HIM,KS}, 
a multifractal analysis of critical wave functions\cite{Terao},  
and plateau transitions in narrow wires\cite{Kawa}.  
In graphene, on the other hand, 
the range of disorder plays an unusually 
important role, since the range dominates the inter-valley 
(K-K') scattering.  Hence we conceive 
here that it is imperative to examine the honeycomb lattice 
(rather than an effective Dirac model) with 
bond randomness of varied correlation lengths in understanding 
the Hall plateau transition in graphene.  

This is exactly our motivation here to explore how the criticality 
in disordered graphene QHE transition, 
especially for $n=0$ Landau level, 
depends on (i) the symmetry and (ii) the range of randomness 
in the honeycomb lattice.  
As for the Hall conductivity which has a topological origin and 
mathematically a Chern number in units of $e^2/h$\cite{TKNN,NTW,Hedge}, 
an unusually accurate 
and efficient method is required for examining 
QHE around the Dirac point (band center) for random systems.  Here 
we have adopted a non-Abelian extension of the Chern-number formalism as 
combined with a lattice-gauge technique.  
We shall show that, while the plateau transition for the 
$n=0$ Landau level has an ordinary critical behavior for 
the {\it un}correlated random bonds, the criticality changes dramatically 
to an anomalous, almost {\it exact fixed point with a 
step-function-like plateau transition} and 
a concomitant delta-function-like Landau level, as soon as 
the spatial correlation in the random bonds exceeds only 
a few bond lengths.  
This can indeed be attributed to the preserved chiral symmetry, 
which is confirmed by adding site randomness to modify the symmetry.

\begin{figure}
\includegraphics[scale=0.3]{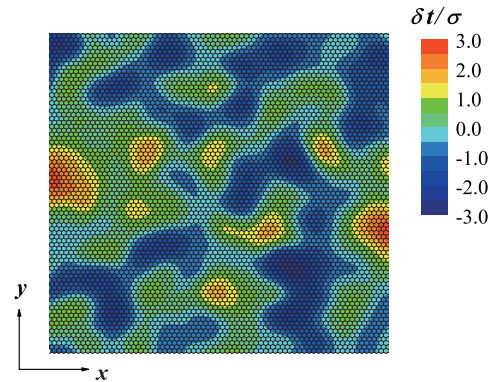}
\caption{(Color online) 
An example of the spatial landscape of the random components, $\delta t/\sigma$, 
in the hopping for a spatial correlation length 
$\eta/a=5$.
\label{fig1}
}
\end{figure}

The tight-binding Hamiltonian for the honeycomb lattice is
$
 H = \sum_{i,j} t_{ij} e^{{\rm i}\theta_{ij}} 
 c_i^{\dagger}c_j$, 
in standard notations, where 
the Peierls phase $\{ \theta_{ij} \}$ is determined such that
the sum of the phases around a hexagon is equal to the magnetic flux 
$-2\pi \phi$ piercing the hexagon in units of the flux 
quantum $\phi_0=h/e$.  
The spin degrees of freedom are neglected for simplicity. 
We introduce randomness in the nearest-neighbor transfer energy as 
$ t_{ij} = t + \delta t_{ij}$, 
where the disordered component $\delta t_{ij}$ is assumed to 
be Gaussian distributed, 
$
 P(\delta t) = 
e^{-\delta t^2/2\sigma^2}/{\sqrt{2\pi \sigma^2}}
$, with a variance $\sigma$.  
Next we specify the spatial correlation $\eta$ in the random components 
by requiring 
$
 \langle \delta t_{ij} \delta t_{kl}
 \rangle = \langle \delta t^2 \rangle e^{-|\mbox{\boldmath $r$}_{ij} -
 \mbox{\boldmath $r$}_{kl}|^2/4\eta^2},
$
where $\mbox{\boldmath $r$}_{ij}$ denotes the 
position of the bond $t_{ij}$, and $\langle \rangle$ the ensemble 
average\cite{Kawa}. We take the $x$- and $y$-axes as shown in Fig.\ref{fig1} 
for $L_x\times L_y$ rectangular systems, where 
a typical spatial landscape of the random hopping is displayed.  
All lengths are measured hereafter in units of the bond length $a$ in 
the honeycomb lattice.  

Even with such random transfers, the Hamiltonian respects the chiral symmetry, 
that is, there exists a local unitary operator $\gamma$ (with $\gamma^2 =1$), which anti-commutes with the Hamiltonian, 
$\{H, \gamma \}=0$.  In real space we can 
decompose the honeycomb lattice into two sub-lattices $A$ and $B$, 
for which the fermion operators are transformed as 
$\gamma c_i \gamma ^{-1}=s c_i$ with $s=+1 (-1)$ for $ i\in A (B)$. 
Obviously, this symmetry is destroyed by a potential disorder, 
while the random hopping preserves it, 
even in magnetic fields.  
Since the eigenstates 
appear in chiral pairs ($\psi, \gamma \psi$ 
with eigenenergies $\pm E$), 
it is clear that
the zero-energy states are special.
If a zero-energy state $\psi$ is not an eigenstate of the chiral operator, 
we can use the zero-energy chiral pairs, $\psi$ and $\gamma \psi$, 
to make them eigenstates of the chiral operator, $\gamma \psi_\pm=\pm \psi_\pm$ with $\psi_\pm=\psi\pm\gamma \psi$.  
Then all the zero modes are eigenstates
of $\gamma $ with amplitudes residing only 
on one of the A and B sub-lattices.
Hence the topologically protected zero-energy 
Landau levels, particularly their criticality, 
can be very sensitive to whether the disorder 
respects  the chiral symmetry or not.

Let us first look at the result for the density of states around 
$E=0$.  
A key interest is how the $n=0$ Landau level is 
broadened by randomness as compared with $n\neq 0$ levels.  
In the calculation of the density of states 
we adopt the Landau gauge 
for the corresponding bricklayer lattice\cite{WFAS} with 
periodic boundaries in $y$-direction and armchair edges in $x$ 
to remove the contribution from zigzag edges.  
The density of states 
$\langle \rho_i \rangle = -\sum_i{\rm Im}G_{ii}(E+{\rm i}
 \gamma)/N\pi$, 
is obtained in terms of the Green's function\cite{SKM}, 
$G_{ii}(E+{\rm i}\epsilon) = \langle i |(E-H+{\rm i}\epsilon)^{-1} | i \rangle$, 
where $N$ is the total number of sites, and $\epsilon$
a small imaginary part in 
energy to evaluate the Green function numerically. 
We have performed the calculation for 
$6.3\times 10^{-4} \leq \epsilon/t \leq 1.0\times 10^{-2}$ and 
confirmed that the anomaly at $E=0$ described below is not affected by the 
value of $\epsilon$.

\begin{figure}
\includegraphics[scale=0.33]{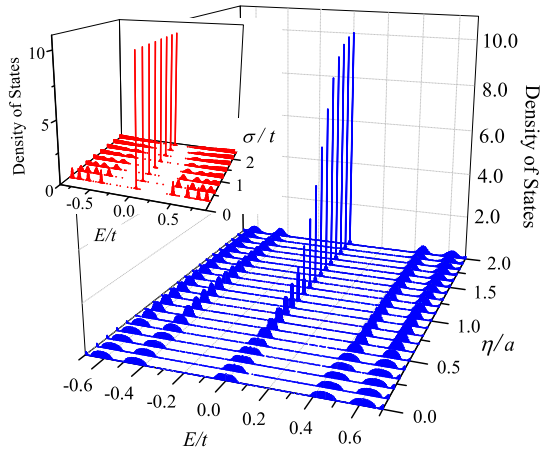}
\caption{(Color online) The density of states for the spatially correlated 
random bonds for various values of the 
correlation length $0\leq \eta/a \leq 2.0$ 
with the strength of disorder fixed at $\sigma/t=0.12$ 
in a magnetic field 
$\phi/\phi_0 = 1/50$. The system size is 
$ L_x/(\sqrt{3}a/2) = 5000, L_y/(3a/2) = 100$, and $\epsilon/t=6.25\times 10^{-4}$.
Inset: The density of states for various values of 
the disorder strength $\sigma$ with a fixed $\eta/a=3$.
\label{fig2}
}
\end{figure}

The result for the density of states with the disorder strength $\sigma/t=0.12$ 
and a magnetic field $\phi/\phi_0 = 1/50$ for 
various values of the correlation length $\eta$ is shown in Fig. \ref{fig2}. 
It has been 
shown\cite{HFA} that the $n=0$ and several adjacent 
Landau levels characteristic to the relativistic electrons are captured 
even with this value of $\phi$, which, when directly 
translated, corresponds to a large magnetic field, 
so the model should be adequate for the analysis of the criticality at the $n=0$ Landau level.
We can immediately see that 
the $n=0$ Landau level is anomalously sharp, but that the sharpness depends 
sensitively on the correlation length $\eta$ of the random hopping. 
More precisely, as soon as we have $\eta /a \geq 1$, 
the $n=0$ Landau level becomes remarkably sharp, 
while this does not occur for $n\neq 0$ Landau levels.  
Indeed, the shape of the $n=0$ Landau level 
for $\eta/a \geq 3$ is delta-function-like within the numerical accuracy 
in that its shape coincides almost exactly with  
the Lorentzian density of states in the clean limit, 
$\rho(E) = \frac{1}{\pi}\frac{\epsilon}{E^2+\epsilon^2}$. 
In this sense the $n=0$ Landau level in the presence of the correlated 
bond randomness is delta-function-like for $\eta/a \geq 3$. 
If we examine the dependence of the 
density of states on the disorder strength, with a fixed $\eta/a =3$ 
(Fig.2, inset), 
we can confirm that this anomaly at $n=0(E=0)$ remains insensitive to the disorder strength as far as 
$\eta/a \gtrsim 1$,  whereas other Landau levels are broadened by disorder.

We now turn to the quantized Hall plateau transition.
The Hall conductivity $\sigma_{xy}$ is related to the Chern
number $n_{\rm C}$ as 
$\sigma_{xy} = n_{\rm C} (e^2/h)$ provided that an energy gap exists 
above the Fermi energy \cite{TKNN,NTW,Hedge}.  
In random systems the Chern number differs from sample 
to sample, so we should look at the ensemble-averaged 
quantity for each energy bin, which gives 
the Hall conductance as a function of $E$ \cite{HIM,AokiAndo96}. 

Since the Hall current is dissipationless, 
not only the state near the Fermi energy but 
all the filled states contribute. 
A speciality of the graphene QHE is that the region of 
interest is around $E=0$, which implies that we have to question 
many Landau levels below the Fermi energy whose contributions 
almost cancel with each other to a value of order unity.  
So we obviously confront a numerically difficult situation, 
especially if we want to look at a criticality around $E=0$. 
We have previously shown that such a situation can be treated with 
a non-Abelian formulation of the Hall conductivity 
as a Chern number for multi-dimensional multiplets of fermions\cite{HC,HFA}. 
The Berry connection is then defined as a 
matrix, which is spanned by the Landau sub-bands, 
split by the randomness.  
We can then adopt an extended unit cell for each realization of the randomness
to apply the formula\cite{NTW}.
The required stable energy gap at the Fermi energy is 
mostly guaranteed by the level repulsion in finite, random systems.  
Level crossings below the Fermi energy do not cause any problem either 
in this formulation.
In the numerical evaluation of the topological numbers, 
a technique developed in the lattice gauge theory has turned 
out to be useful, 
which is a two-dimensional generalization of the
King-Smith-Vanderbilt formula for polarization\cite{FHS,KSV}.
For this we employ twisted boundary conditions, 
$
\psi(x+L_x,y) = e^{{\rm i}\phi_x}\psi(x,y), 
\psi(x,y+L_y) = e^{{\rm i}\phi_y}\psi(x,y), 
$
where $\phi_{x(y)} = 2\pi n_{x(y)}/N$ with $n_{x(y)}=0,1,2,\ldots, N-1$ 
being discretized phases. 
Here the string gauge is used to treat weak magnetic fields 
with the twisted boundary condition\cite{HIM}.  
The Chern number is then evaluated as\cite{HC,FHS,HIM}
$
n_{\rm C} =  \frac{1}{2\pi} \sum_{\bm{\phi}} 
{\rm  arg} ( \det 
\bm{U}_x^{\bm{\phi}}
\bm{U}_y^{\bm{\phi+\Delta \phi_x}}
{[}
\bm{U}_x^{\bm{\phi+\Delta \phi_y}}
\bm{U}_y^{\bm{\phi}} {]}^* ) ,
$
where $\bm{U}_{x(y)}^{\bm{\phi}}= {[}\bm{\Psi} (\bm{\phi} +\Delta \bm{\phi}_{x(y)} ) {]} ^\dagger \bm{\Psi}(\bm{\phi})$
with  a set of eigenstates $\bm{\Psi}(\bm{\phi}) = ( | \psi _1(\bm{\phi} )\rangle ,\cdots, | \psi _M (\bm{\phi} )\rangle)$
below the Fermi energy
and $\Delta \bm{\phi}_x=(2\pi/N,0) $, $\Delta \bm{\phi}_y=(0,2\pi/N) $.

\begin{figure}
\includegraphics[scale=0.3]{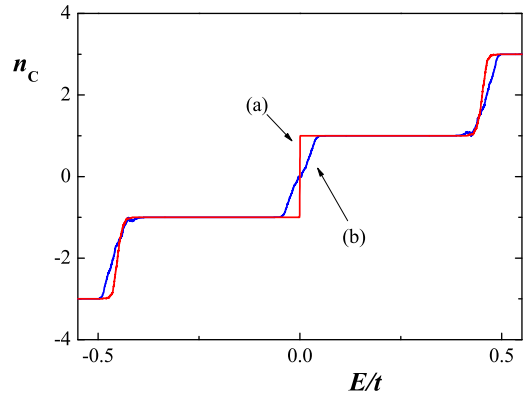}
\caption{(Color online) The Hall conductivity (Chern number) against 
the Fermi energy $E/t$ for spatially correlated random bonds 
for the correlation length $\eta/a = 1.5$(a) and $\eta/a=0$(b). 
We have a disorder strength $\sigma/t=0.12$, a 
magnetic field $\phi/\phi_0 = 1/50$, a system size 
$L_x/(\sqrt{3}a/2)=L_y/(3a/2)=20$, $N=10$ and 
an average over 300 samples.
\label{fig3}
}
\end{figure}

\begin{figure}
\includegraphics[scale=0.3]{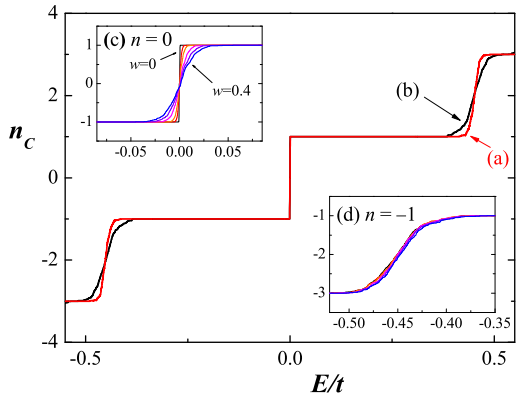}
\caption{(Color online) The Hall conductivity (Chern number) as a function 
of the Fermi energy $E/t$ for spatially correlated 
random bonds with $\eta/a = 1.5$ for 
two system sizes, $L_x/(\sqrt{3}a/2)=L_y/(3a/2)=20$(a) and $=10$(b).  
We have a disorder strength $\sigma/t=0.12$, 
a magnetic field $\phi/\phi_0 = 1/50$ with 
an average over 300 samples.
Insets: Plateau transitions for $n=0$ (c) and $n=-1$ (d) 
when we add a potential disorder 
introduced as random site-energies 
uniformly distributed over $[-w/2,w/2]$ and spatially 
uncorrelated, and the curves are for $w/t=0, 
0.05, 0.1, 0.2, 0.3, 0.4$ with $L_x/(\sqrt{3}a/2)=L_y/(3a/2)=10$. 
\label{fig4}
}
\end{figure}

The Chern number averaged over 300 realizations of randomness 
for a  bond disorder strength $\sigma/t=0.12$ 
is shown in Figs. \ref{fig3} and \ref{fig4} 
plotted within the energy 
 region ($|E/t|<1$), where the plateaus have 
the Dirac behavior, $(2m+1)(e^2/h)$ with 
$m$ an integer\cite{HFA}. 
If we first look at 
Fig. \ref{fig3} for the correlated 
disorder with $\eta/a=1.5$, we immediately notice  
that the plateau transition
between $n_{C}=-1$ and $1$ around $E=0$ is anomalously 
abrupt, i.e., $n_C$ behaves like a step function, in sharp contrast with 
other transitions for $n\neq 0$. 
For the uncorrelated bond randomness  $\eta/a =0$, on the other hand, 
the transition for $n=0$ is as smeared as those for $n \neq 0$. 
The anomalously 
sharp step for $n=0$ agrees with the anomalously sharp $n=0$ Landau 
level seen in the density of states.

We can confirm the behavior for $n=0$ is indeed  unusual by 
looking at a system-size dependence for a 
correlated disorder in Fig. \ref{fig4}, in which 
the results for two sizes coincide with each other within 
the numerical accuracy for the transition at $E=0$, while 
other transitions exhibit the usual behavior of narrower transition 
widths for larger systems.  
We can further test our picture that the anomalous behavior is 
connected to
the preserved chiral symmetry.  For this purpose, 
we have added a potential disorder to the bond disorder (Fig.4, insets).
The result clearly shows that the addition of a 
potential disorder that destroys the chiral symmetry 
does wash out the anomalous (step-function-like) 
transition at $E=0$ into a normal behavior (Fig.4(c)), while other transitions 
remain essentially the same (Fig. 4(d)).

In summary we have revealed that the quantum Hall transition at $E=0$ is 
anomalously sensitive to the spatial correlation of the 
random bonds, where 
concomitantly with 
the Landau level width, 
it 
becomes exact fixed-point-like 
as soon as the correlation length exceeds a few times 
the bond length. 
This sharply contrasts with the case of the generic random Dirac fermions,
where the broadening of the Landau level occurs also for $E=0$.  
The singular behavior may correspond to 
the fixed point for the random Dirac fermions 
with chiral symmetry discussed by Ludwig et al\cite{LFSG},  
where a generic instability of  the fixed point is discussed.  

Experimentally, the length scale of ripples is estimated to be several 
nanometers\cite{Meyer,Fasolino}. Since this is much greater than 
the correlation length adopted here, 
the bond disorder by such ripples should not  
broaden the $n=0$ graphene Landau level.  Conversely, 
the broadening at the $n=0$ level as observed in experiments should be caused by other types of disorder, such as 
potential disorders by charged impurities \cite{Neto}. 
The message here amounts to that 
the fixed-point behavior 
should be experimentally observed in free-standing clean graphene 
samples where the ripple is the only disorder.

\begin{acknowledgments}
We wish to thank Yoshiyuki Ono, Tomi Ohtsuki and Allan MacDonald 
for valuable discussions and comments.
 YH appreciates discussion with B. D. Simons for
 pointing out importance of correlated randomness in Dirac fermions 
after Ref.\protect\cite{HWK} were published.
The work was supported in part by Grants-in-Aid for Scientific Research,
Nos. 20340098 (YH and HA)
 and 20654034 
 from JSPS and
Nos. 220029004 
and 20046002 
on Priority Areas from MEXT for YH. 

\end{acknowledgments}


\vfill
\end{document}